\documentclass[aps,pre,showpacs,superscriptaddress,floatfix,twocolumn]{revtex4-1}
 
\usepackage{graphicx}
\usepackage{color}
\usepackage{braket}
\usepackage{amsmath}
\usepackage{amsthm}   
\usepackage{amsfonts}
\usepackage{amssymb}
\definecolor{darkblue}{rgb}{0,0,.5}
\usepackage[linktocpage, colorlinks=true ,linkcolor=darkblue, citecolor=darkblue]{hyperref}
\usepackage[all]{hypcap}
\usepackage{bm}

\newcommand{\vc}[1]{\boldsymbol{#1}}
\newcommand{\vt}[3]{#1^{#2}_{\phantom{#2}#3}}

\newcommand{\iu}{\mathrm{i}}
\newcommand{\tr}{\mathrm{tr}}

\begin{document}

\title{Interplay of different environments in open quantum systems: Breakdown of the additive approximation}

\author{Giulio~G. \surname{Giusteri}}
\email{giulio.giusteri@oist.jp}
\affiliation{Mathematical Soft Matter Unit, Okinawa Institute of Science and Technology Graduate University, 1919-1 Tancha, Onna, 904-0495, Okinawa, Japan}
\affiliation{Dipartimento di Matematica e
Fisica and Interdisciplinary Laboratories for Advanced Materials Physics,
 Universit\`a Cattolica del Sacro Cuore, via Musei 41, I-25121 Brescia, Italy}
\affiliation{Istituto Nazionale di Fisica Nucleare,  Sezione di Pavia, via Bassi 6, I-27100, Pavia, Italy}
\author{Filippo \surname{Recrosi}}
\affiliation{Gran Sasso Science Institute,
Viale Francesco Crispi 7, I-67100, L'Aquila, Italy}
\author{Gernot \surname{Schaller}}
\affiliation{Institut f\"ur Theoretische Physik, Technische Universit\"at Berlin, Hardenbergstra{\ss}e 36, D-10623 Berlin, Germany}
\author{G.~Luca \surname{Celardo}}
\affiliation{Dipartimento di Matematica e
Fisica and Interdisciplinary Laboratories for Advanced Materials Physics,
 Universit\`a Cattolica del Sacro Cuore, via Musei 41, I-25121 Brescia, Italy}
\affiliation{Istituto Nazionale di Fisica Nucleare,  Sezione di Pavia, via Bassi 6, I-27100, Pavia, Italy}
\affiliation{Benem\'erita Universidad Aut\'onoma de Puebla, Instituto de F\'isica, Apartado Postal J-48, Puebla 72570, Mexico}

\begin{abstract}     
We analyze an open quantum system under the influence of more than one
environment: a dephasing bath and a probability-absorbing bath that
represents a decay channel, as encountered in many models of quantum networks.
In our case, dephasing is modeled by random
fluctuations of the site energies, while the absorbing bath is modeled
with an external lead attached to the system.   
We analyze under which conditions the effects of the two baths can
enter additively the quantum master equation.
When such additivity is legitimate, the reduced master equation corresponds to the evolution generated by an effective non-Hermitian Hamiltonian and a Haken--Strobl dephasing super-operator.
We find that the additive decomposition is a good approximation when the strength of dephasing is small compared to the bandwidth of the probability-absorbing bath.
\end{abstract}                                                               
                                                                            
\date{\today}

\maketitle

%%%%%%%%%%%%%%%%%%%%%%%%%%%%%%%%%%%%%%%%%%%%%%%%%%%%%%%%%%%%%%%
\section{Introduction}

Open quantum systems are nowadays at the center of many research fields 
in physics, ranging from quantum computing to 
transport in nano- and meso-scale solid-state systems as well as biological aggregates.
In particular, charge/excitation transport in the quantum coherent regime
can be considered one of the central subjects in
modern solid-state physics~\cite{Beenakker,Lee,kaplan,dolcini,felix} and in quantum biology~\cite{plenio,cao}. 
When a quantum system interacts with other systems, it is often impossible to treat in detail the full unitary and coherent quantum dynamics of the cumulative structure. 
It is then necessary to restrict attention to a limited portion of it, which is referred to as an open quantum system, while surrounding systems---typically much larger---are called external baths.
Neglecting the detailed evolution of the surrounding has two important consequences on the dynamics of the open quantum system: 
(i) We can have a leakage of excitation from the system. 
(ii) The ignorance of the detailed coherences developed between the system and the baths makes the effective evolution incoherent. 
Typically, these effects are induced by the presence of (i) a decay channel and (ii) a thermal bath.

Open quantum systems in relevant physical situations often interact with more than one environment. 
In the literature there are many examples of systems in which the effects of different environments are treated separately and added as independent terms in the
master equation~\cite{deph,srfmo,mukameldeph,lussardi,alberto}. 
Nevertheless, the fact that two different baths interact with the very
same system would cause them to interact as well~\cite{schaller2016a}.
Consequently, that they affect the system in an independent way is usually true only at the lowest perturbative orders.
It is then very important to understand what is the scope of applicability of the independence hypothesis, 
which is at the basis of so many models proposed in the literature.

{We identify the independence hypothesis with an additive approximation in the following sense.
We assume that the isolated action of each bath on the system can be described in the master equation formalism by a Liouvillian super-operator which, by construction, does not depend on the parameters of any other bath.
Then we consider the Liouvillian super-operator describing the combined action of multiple baths on the system.
The various baths can be considered independent if the collective Liouvillian is well approximated by the sum of the single-bath Liouvillian super-operators.
Our main objective is to investigate conditions under which such additive approximations are legitimate.}

Tight-binding networks provide paradigmatic models, often successfully employed to capture essential physical effects. 
Their coupling with external environments can be taken into account in different ways. 
The action of decay channels (losses by recombination, trapping of the excitation into draining structures, etc.)\ is usually included by adding non-Hermitian terms to the Hamiltonian~\cite{MW,Zannals,rottertb,deph,ZeleReview}.
Other important baths are those inducing static disorder (space-dependent) or dynamical disorder (time-dependent). 
These can be modeled in the framework of quantum master equations in Lindblad form.
Notably, when both disorder and decay channels affect the open system, the strength of the coupling to the decay channel
is usually assumed to be unaffected by the presence of disorder. 
This is a prominent example of the independence hypothesis mentioned above, the scope of which we intend to assess in the present paper.

We discuss this issue by analyzing a simple 
model (Fig.\ \ref{fig:r-l-system}) in which $N_R$ two-level systems (sites) are
arranged in a ring-like structure. Such ring structures are relevant
in natural light-harvesting complexes~\cite{schulten1} and in engineered
devices for light-harvesting and photon sensing~\cite{superabsorb}.
The ring is in contact with a dephasing bath that leads to uncorrelated time-dependent fluctuations of the energy of each site. 
Assuming a white-noise structure of the disturbances, such a bath is treated in the framework of the Haken--Strobl master equation~\cite{HS}.
Furthermore, the ring interacts with a probability-absorbing bath equally coupled to all of its sites. 
A common model for such a decay channel is a one-dimensional lead, that corresponds to a chain of $N_L$ two-level systems in the limit $N_L\to\infty$. 
Similar structures have been used to describe exciton transport in
natural or engineered systems in which the single-excitation approximation is legitimate and
equivalent tight-binding models can be introduced~\cite{lussardi,Jaggr,superabsorb,sarovarbio,mukamelspano,schulten1,fra}.

Based on the tight-binding model described in the following section, we analyze the dynamics of the extended system that includes, 
together with the ring subject to dephasing noise, the linear chain representing the lead.
We derive under which conditions the coupling to the lead and the presence
of the dephasing bath can be treated independently in building the reduced master equation of the sole ring.
Both the analytical derivation presented in Sect.~\ref{sec:reduction} and the numerical results of Sect.~\ref{sec:numerics} 
show that the additive approximation (usually adopted in the literature) is only valid when
the strength of the fluctuations producing dephasing is small compared
to the bandwidth of the probability-absorbing lead. 
{We stress that the system (the ring) is always kept in a regime such that the isolated action of each bath is well-represented by Markovian Liouvillian super-operators.
What we analyze is under which condition the sum of those independent super-operators fails to describe the 
combined effect of both baths on the system.}

Our analysis also confirms that, 
for sufficiently weak noise, the effects of the two baths are independent.
In this case, the master equation is defined by the the sum of the contributions generated by an effective 
non-Hermitian Hamiltonian and by the Haken--Strobl dephasing super-operator.
On the other hand, 
we show that a sufficiently strong
noise leads to a breakdown of both the additive approximation
and the effective non-Hermitian evolution.
These findings complete those of a companion paper~\cite{fra}, in which
the combined effect of static disorder and a decay channel was studied
within the framework of the effective non-Hermitian Hamiltonian approach. 
%

%%%%%%%%%%%%%%%%%%%%%%%%%%%%%%%%%%%%%%%%%%%%%%%%%%%%%%%%%%%%%%%
\section{The Model}\label{sec:model}

\begin{figure}[t]
\centering
\includegraphics[width=0.47\textwidth]{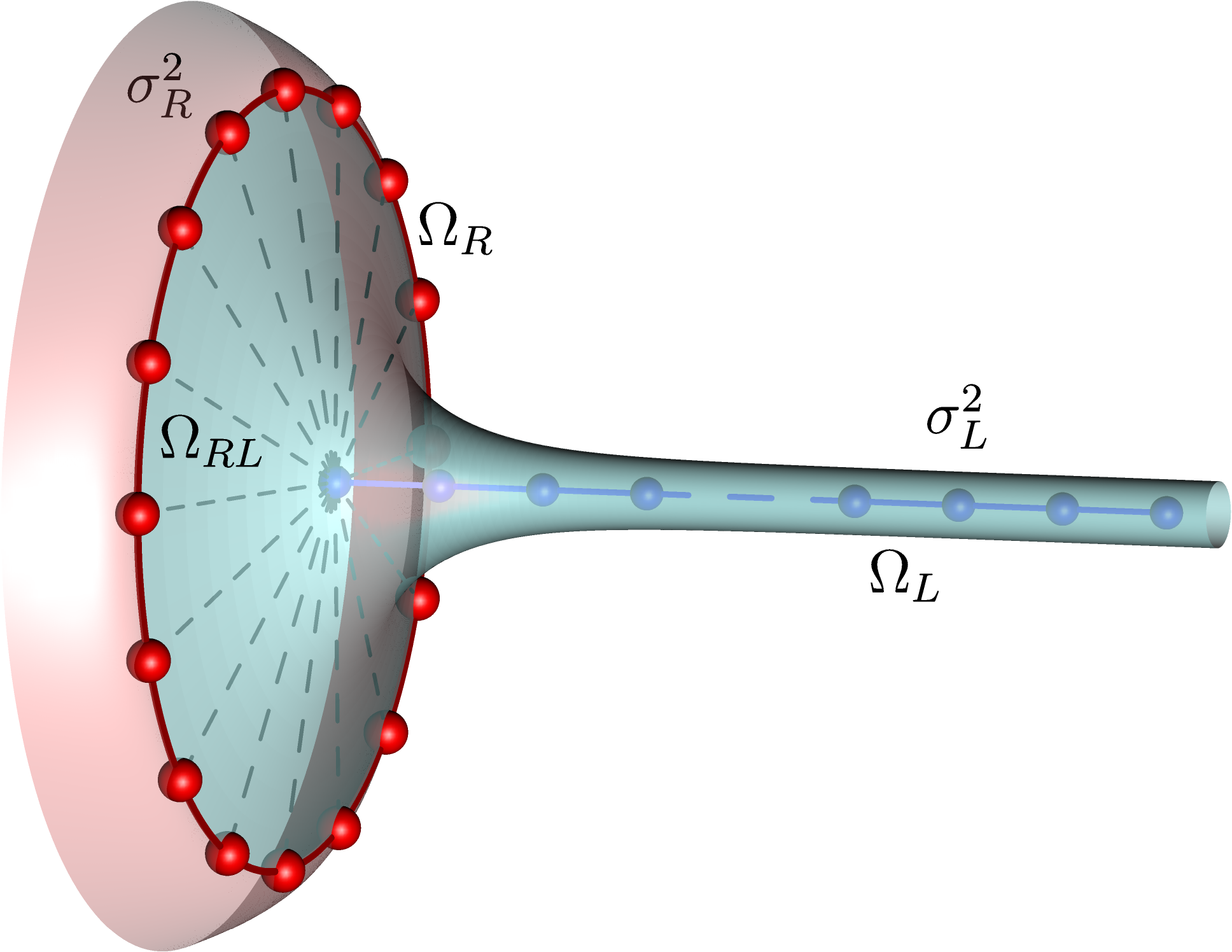}
\caption{(Color Online)\emph{A ring-like network interacts with a dephasing bath (light red) and a probability-absorbing bath (light blue).}
The on-site energies of each of the $N_R$ sites of the ring, connected with nearest-neighbor coupling $\Omega_R$, undergo 
white-noise fluctuations with dephasing strength $\sigma^2_R$. 
The probability-absorbing bath is modeled by a lead of $N_L$ sites, connected with nearest-neighbor coupling $\Omega_L$. 
The ring sites are equally coupled to the first lead site with tunneling amplitude $\Omega_{RL}$ (dashed lines). 
In general, white-noise fluctuations of intensity $\sigma^2_L$ could be present also on the lead (see Appendix~\ref{sec:HS}), 
but only the case $\sigma^2_L=0$ is considered in the main text.}
\label{fig:r-l-system}
\label{rl}
\end{figure} 

We first introduce a Hermitian model to describe the decay of the excitation from the peripheral ring into the chain (Fig.~\ref{rl}). 
The chain represents a probability-absorbing channel, to be considered later in the limit of infinite length.
Specifically, the ring with $N_R$ sites and nearest-neighbor coupling $\Omega_R$ is described by the tight-binding Hamiltonian
\begin{equation}
\label{HR}
H_R = \Omega_R \sum_{\langle r,r'\rangle}\left(|r\rangle \langle r'| + |r'\rangle \langle r|\right)\,, 
\end{equation}
where the sum runs over the pairs of neighboring sites.
Each site of the ring is connected, through the tunneling amplitude $\Omega_{RL}$, to the first site of a lead,
described by a linear chain of $N_L$ resonant sites with nearest-neighbor coupling $\Omega_{L}$. 

The total Hamiltonian of the extended system, written in the site basis
\begin{equation}
\label{eq:sitebasis}
\left\{|r_\mu\rangle,|\ell_\nu\rangle,\mu=1,\ldots,N_R,\nu=1,\ldots,N_L\right\}\,,
\end{equation}
reads 
\begin{equation}
H= H_R + H_{L} + H_{RL} = H_0 + H_{RL} \,.
\label{eq:HRL}
\end{equation}
Here, the Hamiltonian for the lead is
\begin{equation}
H_L= \Omega_L\sum_{\nu=1}^{N_L-1} \left( |\ell_\nu\rangle
\langle \ell_{\nu+1} | + |\ell_{\nu+1}\rangle
\langle \ell_\nu | \right)\,, 
\label{HL}
\end{equation}
and the interaction between the ring and the lead is described by
\begin{equation}
H_{RL}=\Omega_{RL} \sum_{\mu=1}^{N_R} \left( |r_\mu\rangle \langle \ell_1| + 
 |\ell_1 \rangle \langle r_\mu|\right) \,,
\label{VRL}
\end{equation}
where $\Omega_{RL}$ is the coupling between the ring sites and the
first site of the lead.
Note that we limit our considerations to the subspace containing a single excitation in
both ring and lead together, such that not all states will participate in the dynamics.

One can imagine that, when $N_L$ is large enough, the lead represents a good sink, in that it absorbs most of the excitation present in the system. 
In reality, the structure of the coupling between the ring and the
lead is such that the decay of the excitation is strongly dependent on
the initial state. 
The symmetry of the extended system leads to the situation that only one of the $N_R$ ring eigenstates is coupled to the lead, 
with a coupling enhanced by a factor of $\sqrt{N_R}$ compared to the single-site coupling~\cite{fra}. 
The super-transferring state $|S\rangle$ is fully symmetric in the site basis of the ring and given by
\begin{equation}
|S\rangle =\frac{1}{\sqrt{N_R}} \sum_{\mu=1}^{N_R} |r_\mu\rangle\,.
\label{SR}
\end{equation}
This completely symmetric superposition of site states will decay with an enhanced rate, proportional to $N_R$, 
giving rise to the phenomenon of superradiance. 
In contrast, the presence of static disorder or noise destroys the symmetry, restoring a democratic coupling 
of the ring states with the lead, and generating an overall decay rate
independent of the system size $N_R$ \cite{lussardi,CelGiuBor14}. 
Such superradiant effect is a specific feature of this model but it is
not essential to the results presented here. 

{A more important fact, also pointed out in Ref.~\cite{fra}, is that the lead can be effectively represented as a decay channel only if the coupling $\Omega_R$ between ring sites is small compared to the lead bandwidth, determined by the coupling $\Omega_L$.
We thus assume $\Omega_R\ll\Omega_L$ since it is only in this regime that the lead can be seen as a probability-absorbing bath for the ring system.
We therefore neglect the ring coupling in the following analytical treatment.
However, we confirm through numerical simulations that a finite but small value
of $\Omega_R$ does not affect our results, see
Fig.~\ref{fig:decay-rates} and the related discussion.}

We now introduce a dephasing bath by assuming the presence of white-noise fluctuations on the excitation energy of the system sites.
This means that the energies $\varepsilon^R_\mu=\hbar q^R_\mu$ of the ring sites undergo independent white-noise fluctuations with intensity 
$\sigma^2_R$, i.e., formally the frequencies $q^R_\mu$ satisfy the relation
\begin{equation}
\langle q^R_\mu(t)q^R_\nu(t')\rangle=\frac{\sigma^2_R}{\hbar}\delta_{\mu\nu}\delta(t-t')\,.
\end{equation}
We thus identify the energy scale $\sigma_R^2$ as the dephasing strength on the ring.
We could in principle apply our treatment also with the site energies of the lead that fluctuate with intensity $\sigma^2_L$, 
but this would lead to a direct coupling between the dephasing and the dissipative bath (lead), 
thereby obscuring the main effect that we want to analyze.
{We thus set $\sigma^2_L=0$ in what follows.}

The quantum master equation that describes the evolution of the density matrix in the presence of such a dephasing noise is the 
Haken--Strobl~\cite{HS} equation. 
Here, we briefly recall its form, but in Appendix~\ref{sec:HS} we present a simple derivation of this result 
(in which also the energies of the lead sites can fluctuate), obtained by exploiting It\^o's stochastic calculus.

It is now convenient to view our network as a bipartite system. 
We thus label the states in the single-excitation subspace of the total Hilbert space as
\begin{equation}
\ket{i}\ket{0_L}\qquad(i=1,\ldots,N_R)\,,
\end{equation}
if the single excitation is on the $i$-th ring site, and
\begin{equation}
\ket{0_R}\ket{i}\qquad(i>N_R)\,,
\end{equation}
if the single excitation is on the $(i-N_R)$-th lead site.
With $\ket{0_R}$ and $\ket{0_L}$ we denote the vacuum state on the ring and on the lead, respectively.

With this notation, the ring-lead density matrix admits the following representation:
\begin{equation}\label{eq:rldensity}
\begin{aligned}
\rho(t)=\mbox{}&\sum_{i,k>N_R}c_{i} c^{*}_{k}\ket{0_R}\bra{0_R}\otimes\ket{i}\bra{k}\\
&\mbox{}+\sum_{i,k\leq N_R}c_{i} c^{*}_{k}\ket{i}\bra{k}\otimes\ket{0_L}\bra{0_L}\\
&+\sum_{i\leq N_R,k>N_R}c_{i}c^{*}_{k}\ket{i}\bra{0_R}\otimes\ket{0_L}\bra{k}\\
&+\sum_{i\leq N_R,k>N_R}c^{*}_{i}c_{k}\ket{0_R}\bra{i}\otimes\ket{k}\bra{0_L}\,.
\end{aligned}
\end{equation}

The Haken--Strobl equation for the components $\rho_{ik}=c_ic_k^*$ of the density matrix of the ring-lead system in the single-excitation subspace reads
\begin{equation}
\dot{\rho}_{ik}=-\frac{\iu}{\hbar}([H_0+H_{RL},\rho])_{ik}-(1-\delta_{ik})\frac{\sigma^2_{ik}}{\hbar}\rho_{ik}\,,
\label{eq:rfull}
\end{equation}
where no summation on repeated indices is assumed, and
\begin{equation}\label{eq:HS-rl}
\sigma^2_{ik}=\left\{
\begin{aligned}
\sigma_R^2\qquad&\text{if }i,k\leq N_R\,,\\
\frac{1}{2}\sigma_R^2\qquad&\text{if }i\leq N_R,k>N_R\,,\\
0\qquad&\text{if }i,k> N_R\,.
\end{aligned}\right.
\end{equation}
We present in Appendix~\ref{app:LinHS} the corresponding more common Lindblad form of Eq.~\eqref{eq:rfull}, which is not 
restricted to the single-excitation subspace considered in our analysis.

%%%%%%%%%%%%%%%%%%%%%%%%%%%%%%%%%%%%%%%%%%%%%%%%%%%%%%%%%%%%%%%%%
\section{Reduction to the sole ring}\label{sec:reduction}

In our model, the lead represents a probability-absorbing bath. Under the assumptions discussed in the previous section, it is possible to reduce Eq.~\eqref{eq:HS-rl} to a master equation for the sole ring system, representing the combined effects of the dephasing and probability-absorbing baths.

\subsection{Super-operator representation}

To facilitate calculations, we now introduce some super-operators, defined by their action on $\rho$ as follows:
\begin{equation}
\mathcal{H}_0\rho= \frac{\iu}{\hbar}[H_0,\rho]\,,\qquad\mathcal{V}_{RL}\rho=\frac{\iu}{\hbar}[H_{RL},\rho]\,,
\end{equation}
\begin{equation}
[\mathcal{D}_0\rho]_{ik}=\left\{
\begin{aligned}
(1-\delta_{ik})\frac{\sigma_R^2}{\hbar}\rho_{ik}&\text{ if }i,\,k\leq N_R\,,\\
0\qquad&\text{ otherwise},
\end{aligned}\right.
\end{equation}
\begin{equation}
[\mathcal{D}_{RL}\rho]_{ik}=\left\{
\begin{aligned}
\frac{\sigma_R^2}{2\hbar}\rho_{ik}&\text{ if }i(k)\leq N_R,\,k(i)>N_R,\\
0\qquad&\text{ otherwise}.
\end{aligned}\right.
\end{equation}

The master equation in super-operator form reads now
\begin{equation}\label{eq:HSsuper}
\dot{\rho}=-(\mathcal{H}_0+\mathcal{D}_0)\rho-(\mathcal{V}_{RL}+\mathcal{D}_{RL})\rho\,,
\end{equation}
where $(\mathcal{H}_0+\mathcal{D}_0)$ is the non-interacting super-operator. 
The interaction super-operators are $\mathcal{V}_{RL}$, proportional to the coupling $\Omega_{RL}$, 
and $\mathcal{D}_{RL}$, due to the noise terms. 
Whereas $\mathcal{V}_{RL}$ corresponds to a physical interaction, $\mathcal{D}_{RL}$ is of rather informational nature, since it describes the suppression of coherences between ring and lead sites.

Being interested in studying the decay of the excitation from the ring into the lead,
we assume the lead to be in the vacuum state throughout. 
This, together with the usual Born approximation, yields the following form for the total density matrix:
\begin{equation}
\begin{aligned}
\rho(t)&\mbox{}=\rho_R(t)\otimes\ket{0_L}\bra{0_L}\\
&\mbox{}=\sum_{i,k\leq N_R}c_{i}(t) c^{*}_{k}(t)\ket{i}\bra{k}\otimes\ket{0_L}\bra{0_L}\,.
\end{aligned}
\end{equation}

\subsection{Extended interaction picture}

We now move to the interaction picture in the super-operator representation.
We define
\begin{equation}
\rho^I(t)=e^{(\mathcal{H}_0+\mathcal{D}_0)t}\rho(t)
\end{equation}
and
\begin{equation}
\mathcal{V}_{RL}(t)+\mathcal{D}_{RL}(t)=e^{(\mathcal{H}_0+\mathcal{D}_0)t}(\mathcal{V}_{RL}+\mathcal{D}_{RL})e^{-(\mathcal{H}_0+\mathcal{D}_0)t}
\end{equation}
and rewrite equation \eqref{eq:HSsuper} as
\begin{equation}\label{eq:HSsuper-inter}
\dot{\rho}^I=-(\mathcal{V}_{RL}(t)+\mathcal{D}_{RL}(t))\rho^I\,.
\end{equation}
A crucial observation is now that, under the assumption $\rho^I(t)=\rho^I_R(t)\otimes\ket{0_L}\bra{0_L}$, we have $\mathcal{D}_{RL}(t)\rho^I(t)=0$. Consequently, equation \eqref{eq:HSsuper-inter} reduces to
\begin{equation}\label{eq:HSsuper-final}
\dot{\rho}^I=-\mathcal{V}_{RL}(t)\rho^I\,.
\end{equation}

To find the reduced master equation for the ring system, 
we formally solve \eqref{eq:HSsuper-final} and insert the solution into the r.h.s.\ of \eqref{eq:HSsuper-final}, leading to
\begin{equation}\label{eq:traced}
\begin{aligned}
\dot{\rho}^I_R=\mbox{}&-\tr_L\{\mathcal{V}_{RL}(t)\rho^I(0)\}\\
&\mbox{}+\int_0^t\tr_L\{\mathcal{V}_{RL}(t)\mathcal{V}_{RL}(t')\rho^I(t')\}\,dt'\,.
\end{aligned}
\end{equation}
As usual, the first term on the right-hand side of Eq.~\eqref{eq:traced} vanishes, since the non-interacting evolution operator $e^{(\mathcal{H}_0+\mathcal{D}_0)t}$ annihilates the vacuum state on the lead, present in the initial condition $\rho^I(0)$.

We will now compute an explicit expression for the foregoing equation in the case $\Omega_R=0$ (no hopping on the ring), in which we can diagonalize the non-interacting super-operator $\mathcal{H}_0+\mathcal{D}_0$ on the basis
\begin{equation}\label{eq:basis}
\left\{
\vc{\alpha}_{ik}=\ket{\alpha_i}\bra{\alpha_k}:i,k=1,\ldots,N_R+N_L
\right\}\,,
\end{equation}
where, denoting by $\ket{E_k}$ the lead eigenstates, $\ket{\alpha_k}=\ket{k}\ket{0_L}$ for $k\leq N_R$ and $\ket{\alpha_k}=\ket{0_R}\ket{E_k}$ for $k> N_R$. We will denote the eigenvalue of the non-interacting super-operator $\mathcal{H}_0+\mathcal{D}_0$ associated with $\vc{\alpha}_{ik}$ by $\alpha_{ik}$. Clearly, the populations of the lead eigenstates are not evolving in time under the noninteracting super-operator. Consequently, $\alpha_{kk}=0$ for $k>N_R$.

The expression of $\mathcal V_{RL}(0)$ on such basis and in the continuum limit $N_L\to\infty$ is given by
\begin{equation}
\begin{aligned}
&\mathcal V_{RL}(0)\rho\mbox{}=
\frac{\iu}{\sqrt{N_R}}\sum_{k=N_R+1}^{N_R+N_L}\sum_{i=1}^{N_R}g^*_{E_k}\Big[
\vc{\alpha}_{ik}+\vc{\alpha}_{ki},\,\rho\,\Big]\\
&\mbox{}={\iu}\int d{E}f(E)\sum_{i}\frac{g^*_{E}}{\sqrt{N_R}}\Big[
\ket{i}\ket{0_L}\bra{0_R}\bra{E}+\mathrm{h.c.},\,\rho\,\Big]\,,
\end{aligned}
\end{equation}
where the sum over $i$ is on the ring sites, the integral is on the lead energies with spectral density $f(E)$, 
and we have
\begin{equation}
\frac{g^*_{E}}{\sqrt{N_R}}=
\frac{\Omega_{RL}\sqrt{2}}{\hbar}\sqrt{1-\left(\frac{E}{2\Omega_{L}}\right)^{2}}\,.
\end{equation}

If we denote by $\mathcal V_{ik,rs}$ the components of $\mathcal{V}_{RL}(0)$ in the basis \eqref{eq:basis}, we can write
\begin{equation}
\begin{aligned}
\int_0^t&\mathcal{V}_{RL}(t)\mathcal{V}_{RL}(t')
\rho^I_R(t')\otimes\ket{0_L}\bra{0_L}\,dt'
=\int_0^t\sum_{ik}\vc{\alpha}_{ik}\times\\
&\mbox{}\times\left(\sum_{lm}\sum_{rs}e^{\alpha_{ik}t}\mathcal V_{ik,lm}e^{-\alpha_{lm}(t-t')}\mathcal V_{lm,rs}\rho_{rs}(t')\,dt'\right).
\end{aligned}
\end{equation}
In the previous expression, the operators $\vc{\alpha}_{ik}$ are the elements of the basis introduced in Eq.\ \eqref{eq:basis}, with eigenvalues $\alpha_{ik}$.
Due to our assumption on the density matrix, the sum over $rs$ comprises only ring components, that is $r,s=1,\ldots,N_R$.
Now, $\mathcal V_{lm,rs}$ vanishes if we have either $l\leq N_R$ and $m\leq N_R$ or $l> N_R$ and $m> N_R$.

\subsection{Trace over the lead}

By taking the partial trace over the lead bath we want to find a reduced super-operator that acts only on the reduced density matrix of the ring.
This can be expressed in the basis
\begin{equation}\label{eq:basis-reduced}
\left\{
\vc{r}_{ik}=\ket{i}\bra{k}:i,k=0,1,\ldots,N_R
\right\}\,,
\end{equation}
where we have introduced also the ring vacuum population and the related coherences.

\begin{widetext}
Making the limit $N_L\to\infty$ explicit and recalling that $\alpha_{EE}=0$, we obtain
\begin{equation}
\begin{aligned}
\tr_L&\left\{\int_0^t\mathcal{V}_{RL}(t)\mathcal{V}_{RL}(t')
\rho^I_R(t')\otimes\ket{0_L}\bra{0_L}\,dt'\right\}=\\
=\mbox{}&\int_0^t\sum_{i,k=1}^{N_R}\vc{r}_{ik}\left(\int dE'\,f(E')\sum_{m,r,s=1}^{N_R}e^{(\alpha_{ik}-\alpha_{E'm})t}\mathcal V_{ik,E'm}\mathcal V_{E'm,rs}\rho_{rs}(t')e^{\alpha_{E'm}t'}\right)\,dt'\\
&\mbox{}+\int_0^t\sum_{i,k=1}^{N_R}\vc{r}_{ik}\left(\int dE'\,f(E')\sum_{l,r,s=1}^{N_R}e^{(\alpha_{ik}-\alpha_{lE'})t}\mathcal V_{ik,lE'}\mathcal V_{lE',rs}\rho_{rs}(t')e^{\alpha_{lE'}t'}\right)\,dt'\\
&\mbox{}+\int_0^t\vc{r}_{00}\int d{E}\left(\int dE'\,f(E')\sum_{m,r,s=1}^{N_R}e^{-\alpha_{E'm}t}\mathcal V_{EE,E'm}\mathcal V_{E'm,rs}\rho_{rs}(t')e^{\alpha_{E'm}t'}\right)\,dt'\\
&\mbox{}+\int_0^t\vc{r}_{00}\int d{E}\left(\int dE'\,f(E')\sum_{l,r,s=1}^{N_R}e^{-\alpha_{lE'}t}\mathcal V_{EE,lE'}\mathcal V_{lE',rs}\rho_{rs}(t')e^{\alpha_{lE'}t'}\right)\,dt'\,.
\end{aligned}
\end{equation}
\end{widetext}
In the previous expression, the operator terms are given by the elements $\vc{r}_{ik}$ of the basis defined in Eq.~\eqref{eq:basis-reduced}.

We now substitute the expressions
\begin{equation}
\alpha_{mE}=-\frac{\iu}{\hbar}E=-\alpha_{Em}\,,
\quad\alpha_{rs}={\sigma_R^2}(1-\delta_{rs})\,,\
\end{equation}
\begin{equation}
\mathcal V_{E'm,rs}=\frac{{\iu}g^*_{E'}}{\sqrt{N_R}}\delta_{ms}\,,
\quad\mathcal V_{lE',rs}=-\frac{{\iu}g^*_{E'}}{\sqrt{N_R}}\delta_{lr}\,,
\end{equation}
\begin{equation}
\mathcal V_{ik,E'm}=\frac{{\iu}g^*_{E'}}{\sqrt{N_R}}\delta_{km}\,,
\quad\mathcal V_{ik,lE'}=-\frac{{\iu}g^*_{E'}}{\sqrt{N_R}}\delta_{il}\,,
\end{equation}
\begin{equation}
\mathcal V_{EE,E'm}=-\frac{{\iu}g^*_{E'}}{\sqrt{N_R}}\delta(E-E')=-
\mathcal V_{EE,mE'}\,,
\end{equation}
and
\begin{equation}
\begin{aligned}
&J(E') = |g^*_{E'}|^2 f(E')=\\
&=
\begin{cases}
\frac{\Omega_{RL}^2}{\pi \hbar^2 \Omega_L} \sqrt{1-\left(\frac{E'}{2\Omega_L}\right)^2}
& \text{for} \, E \in [-2\Omega_L, \, 2\Omega_L] \,,\\
0 & \text{otherwise.}
\end{cases}
\end{aligned}
\end{equation}
Then, the partial trace becomes
\begin{equation}\label{eq:finite-band-ME}
\begin{aligned}
\tr_L&\left\{\int_0^t\mathcal{V}_{RL}(t)\mathcal{V}_{RL}(t')
\rho^I_R(t')\otimes\ket{0_L}\bra{0_L}\,dt'\right\}=\\
=\mbox{}&-\int_0^tdt'\sum_{i,k=1}^{N_R}\vc{r}_{ik}\int dE'\,J(E')e^{\alpha_{ik}t}\times\\
&\times\sum_{r,s=1}^{N_R}\left(e^{-\frac{\iu}{\hbar}E'(t-t')}\delta_{ks}\rho_{rs}(t')+e^{\frac{\iu}{\hbar}E'(t-t')}\delta_{ir}\rho_{rs}(t')\right)\\
&\mbox{}+2\int_0^tdt'\vc{r}_{00}\int dE'\,J(E')\sum_{r,s=1}^{N_R}\rho_{rs}(t')\cos\frac{E'(t-t')}{\hbar}\,.
\end{aligned}
\end{equation}

Substituting Eq.~\eqref{eq:finite-band-ME} into Eq.~\eqref{eq:traced} would still entail a term that is non-local in time. To reach a local form of the reduced master equation we need further approximations. 

\subsection{Wide-band limit}

To understand better what are the crucial approximations, we first simplify the kernel $J(E')$ by setting $J(E')\equiv J(0)$ 
for $E'\in[-2\Omega_L,2\Omega_L]$, and zero otherwise. 
Such an approximation preserves the bandwidth of the decay channel while changing the profile of the density of states. 
Since this change is negligible close to the center of the band, it is expected to be a good approximation when the 
ring energies lie close to center of the lead energy band. 
Moreover, it has been noted multiple times (see, for instance, Refs.~\cite{peres,pastawski,fra}) that the profile of 
the density of states close to the edges of the band influences the long-time behavior of the decay, but not its initial features.

Then, we perform the integration over $E'$ in Eq.~\eqref{eq:finite-band-ME} to obtain
\begin{equation}
\begin{aligned}
\tr_L&\left\{\int_0^t\mathcal{V}_{RL}(t)\mathcal{V}_{RL}(t')
\rho^I_R(t')\otimes\ket{0_L}\bra{0_L}\,dt'\right\}=\\
=&\mbox{}-\int_0^t\sum_{i,k=1}^{N_R}dt'\vc{r}_{ik}\frac{2\pi \hbar J(0)\sin(2\Omega_L(t-t')/{\hbar})}{\pi(t-t')}\times\\
&\mbox{}\times\sum_{r,s=1}^{N_R}e^{\alpha_{ik}t}(\delta_{ks}+\delta_{ir})\rho_{rs}(t')\\
&\mbox{}+\int_0^tdt'\vc{r}_{00}\frac{4\pi \hbar J(0)\sin(2\Omega_L(t-t')/{\hbar})}{\pi(t-t')}\sum_{r,s=1}^{N_R}\rho_{rs}(t')\,.
\end{aligned}
\end{equation}
We consider the characteristic time of the ring dynamics given by $\hbar/\sigma^2_R$ and introduce the dimensionless interval $\tau=\sigma^2_R(t-t')/\hbar$.
Since 
\begin{equation}
\lim_{\omega\to\infty}\frac{\sin(\omega\tau)}{\pi\tau}=\delta(\tau)
\end{equation}
in the sense of distributions, we can obtain a local-in-time equation by substituting $\tau$ in the previous 
expression and taking the wide-band limit $\Omega_L/\sigma_R^2\to\infty$. 

We remark that the wide-band limit is not performed with respect to the energy scale of the ring, which is always assumed negligible compared to $\Omega_L$ in our argument.
What we are comparing here is the bandwidth of the probability-absorbing bath with the energy scale of the dephasing bath.
This operation is responsible for removing back-action effects between the two baths and yields
\begin{equation}\label{eq:wide-band-ME}
\begin{aligned}
\tr_L&\left\{\int_0^t\mathcal{V}_{RL}(t)\mathcal{V}_{RL}(t')
\rho^I_R(t')\otimes\ket{0_L}\bra{0_L}\,dt'\right\}=\\
=&\mbox{}-\sum_{i,k=1}^{N_R}\vc{r}_{ik}\left(\pi \hbar J(0)\sum_{r=1}^{N_R}e^{\alpha_{ik}t}[\rho_{ir}(t)+\rho_{rk}(t)]\right)\\
&\mbox{}+\vc{r}_{00}\left(2\pi \hbar J(0)\sum_{r,s=1}^{N_R}\rho_{rs}(t)\right)\,.
\end{aligned}
\end{equation}

\subsection{Reduced master equation and effective Hamiltonian}

If we now define
\begin{equation}
{\gamma=2\pi\hbar^2 J(0)=2\Omega_{RL}^2/\Omega_L}
\end{equation}
and the decay operator $W$ with matrix elements
\begin{equation}
W_{ik}=\gamma/2
\end{equation}
for $i,k=1,\ldots,N_R$, we can substitute Eq.~\eqref{eq:wide-band-ME} into Eq.~\eqref{eq:traced}, transform 
back to the Schr\"odinger picture and obtain the following equations for the elements of the reduced density matrix
\begin{gather}
\dot{\rho}^R_{00}={\frac{\gamma}{\hbar}}\sum_{r,s=1}^{N_R}\rho^R_{rs}\,,\label{eq:r00}\\
\dot{\rho}^R_{ik}=-\frac{\iu}{\hbar}([H_R,\rho^R]-\iu\{W,\rho^R\})_{ik}-(1-\delta_{ik})\frac{\sigma^2_{R}}{\hbar}\rho^R_{ik}\,.\label{eq:rik}
\end{gather}

Within this approximation, which is good for $\sigma^2_R/\Omega_L\to 0$, the terms encoding the effect of dephasing (proportional to $\sigma_R^2$) and the decay of the excitation (proportional to $\gamma$) enter 
additively in the final master equation \eqref{eq:r00}--\eqref{eq:rik}. 
Retaining higher-order terms in the ratio $\sigma^2_R/\Omega_L$ would necessarily bring in terms involving 
products of $\gamma$ and $\sigma_R^2$. 

It should be noted that, in the absence of dephasing ($\sigma_R^2=0$), Eq.~\eqref{eq:rik} corresponds to the coherent 
evolution on the ring described by the effective non-Hermitian Hamiltonian~\cite{Zannals,fra}
\begin{equation}
H_\mathrm{eff}=H_R-\iu W\,.
\end{equation}
Consequently, we can say that, {when the bandwidth of the decay channel is large compared to the intensity of the noise}, the decay
effects encoded in the non-Hermitian Hamiltonian and the dephasing
effects described by the Haken--Strobl super-operator can be independently
added to the closed-system Hamiltonian $H_R$. 
This is the standard
form found in the literature on excitonic transport~\cite{deph}.

Note that these results have been obtained by setting $\Omega_R=0$, thus neglecting the effects of the coupling between ring sites. Nevertheless, on the basis of the analysis presented in Ref.~\cite{fra}, we expect the present results to remain valid provided that $\Omega_R$ is well within the energy band of the lead.
{This expectation is confirmed by the numerical results presented in Fig.~\ref{fig:decay-rates}.}

\section{Numerical results}\label{sec:numerics}

From the results of the previous sections, we expect that the strength of dephasing $\sigma_R^2$ 
leading to a breakdown of the additive approximation is proportional to the energy 
bandwidth $4\Omega_L$ in the lead.
To confirm this and to obtain an estimate of the actual proportionality factor, we performed some numerical simulations.

We compared the evolution generated by the Haken--Strobl master
equation for the extended system comprising the ring and the lead sites
(see Eq.~\eqref{eq:rfull}) with the evolution generated on the ring (reduced model, see Eqs.~\eqref{eq:r00} and \eqref{eq:rik}) by the additive combination
of the Haken--Strobl terms and the non-Hermitian terms describing the
lead as a decay channel. 

\begin{figure}
\centering
\includegraphics[width=0.47\textwidth]{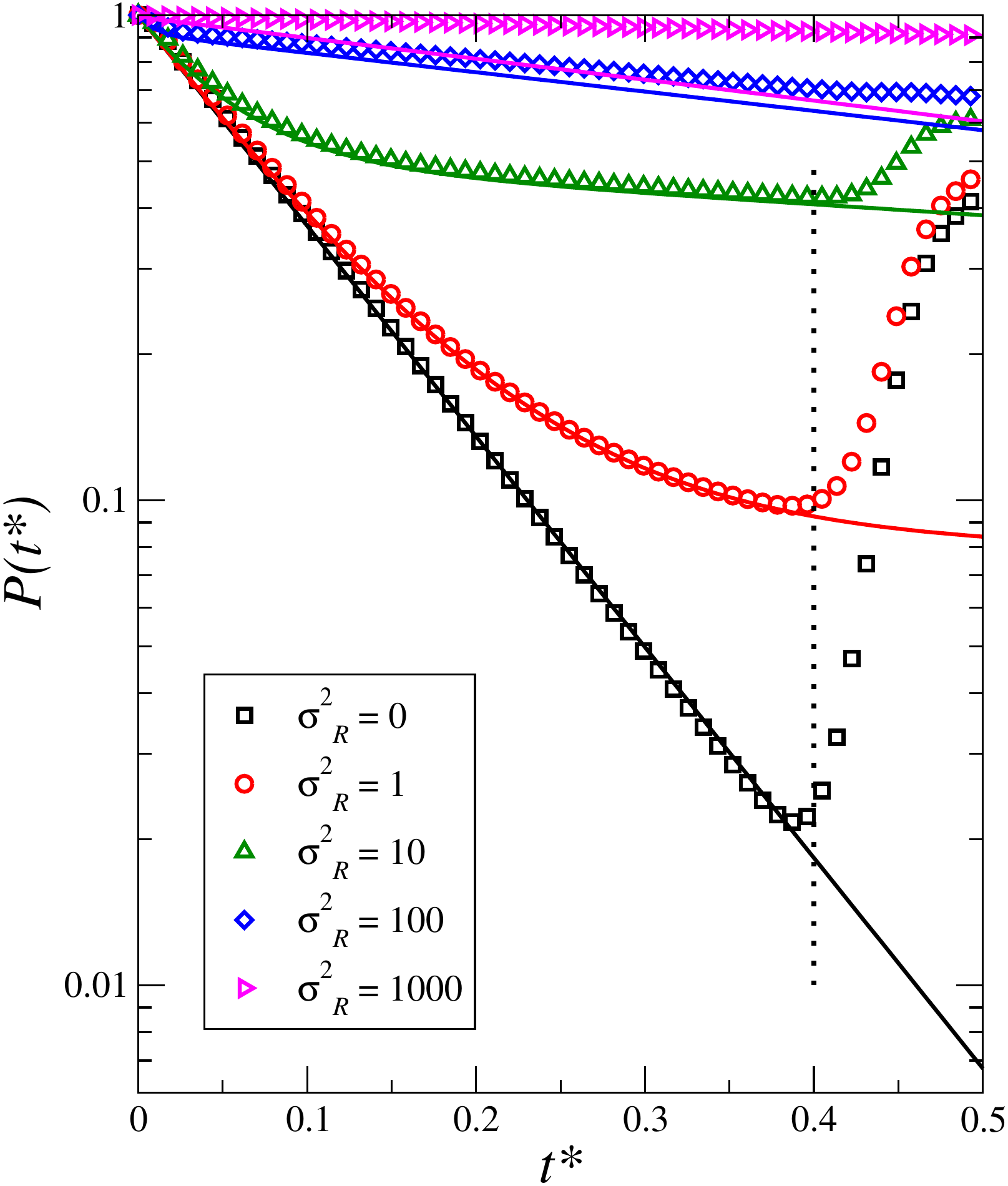}
\caption{(Color Online)\emph{The presence of dephasing noise on the ring first destroys superradiance and then 
leads to the breakdown of the additive approximation}. 
By considering the evolution of the ring population $P(t^*)$, 
in terms of the rescaled time $t^*=\gamma t/\hbar$, we observe that, 
for dephasing strengths smaller than the coupling $\Omega_L$, the evolution of the reduced model 
(curves) agrees with that of the extended model (symbols) up to the insurgence of numerical 
finite-size effects (vertical dotted line), see also the discussion in Ref.~\cite{fra}.
For dephasing strengths larger than $\Omega_L$, there is no agreement between reduced and extended model 
(the decay of solid lines is markedly faster than that of symbols).
Employed values: $\Omega_R=0$, $N_R=10$, ${\gamma=2\Omega_{RL}^2/\Omega_L}=1$, $\Omega_L=100$, $N_L=40$.}\label{fig:agreement}
\end{figure}

First, we studied the probability $P(t)$ of finding the excitation in
the ring at time $t$, giving as initial condition a completely
symmetric superposition of ring sites, see Eq.~\eqref{SR} and Fig.~\ref{fig:agreement}. 
As we already mentioned, in the absence of disorder and noise, such a
superposition is a superradiant state. 
Indeed, its decay width for
$\sigma^2_R=0$ is $N_R$ times larger than the single-site decay width
${\gamma=2\Omega_{RL}^2/\Omega_L}$. 
In the absence of noise, the agreement between the extended
model and the reduced one is excellent, up to a time in which the
finite length of the computational lead produces a spurious revival in
the probability $P(t)$, see vertical dotted line in
Fig.~\ref{fig:agreement} and discussion in Ref.~\cite{fra}. 
Note that as the size of the lead increases to infinity, also the
revival time diverges. 

The agreement persists up to dephasing strengths of the order of the
inter-site coupling $\Omega_L$  within the lead. 
For larger dephasing strength, the extended model features a much slower decay than the reduced model 
(in which the decay width converges to the single-site value $\gamma$)
and the agreement is lost since the very early stages of the
evolution. Indeed, large fluctuations of the ring site energies bring
the energy of the ring outside the energy band of the lead. This
induces a strong suppression of the decay not captured by
the additive approximation, which would predict a finite decay rate, equal to
$\gamma$, also in the limit of infinite noise strength.

\begin{figure}
\centering
\includegraphics[width=0.47\textwidth]{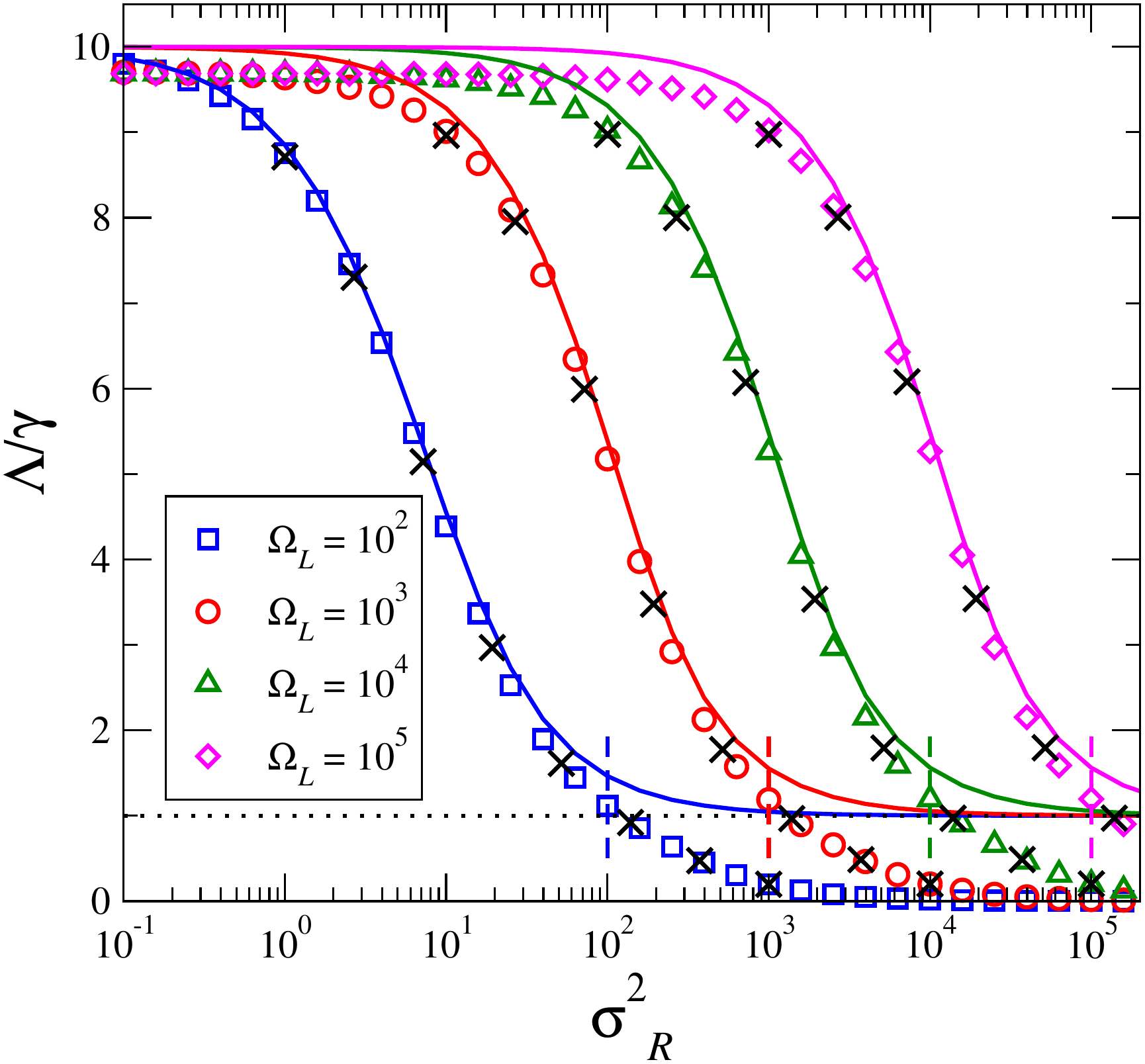}
\caption{(Color Online)\emph{When the dephasing strength $\sigma_R^2$ equals the lead coupling $\Omega_L$, the agreement between reduced and extended model is completely lost}. 
The decay width $\Lambda$ of an excitation initialized on the superradiant state, normalized by the single-site decay 
width $\gamma$, is plotted as a function of the dephasing rate $\sigma^2_R$ affecting the ring for different
values of the coupling $\Omega_L$ (see legend). 
Symbols correspond to the extended model, while solid curves are obtained by the reduced model. 
The agreement is lost when $\sigma_R^2\approx\Omega_L$ (vertical dashed lines), after which point the decay 
rates of the extended model vanish, while those of the reduced model converge to the single-site decay (horizontal dotted line).
Employed values: $\Omega_R=0$, $N_R=10$, $\gamma=1$, $N_L=40$.
{In addition, data obtained by setting $\Omega_R=0.1\Omega_L$ (black crosses) show that a coupling among ring sites small compared to $\Omega_L$ has a negligible effect on the behavior of the system.}}\label{fig:decay-rates}
\end{figure}

To obtain an estimate of the dephasing strength that destroys the
agreement between reduced and extended model, we considered the effective
decay width $\Lambda$ for the superradiant state (extracted by the
curves of $P(t)$) as a function of the dephasing strength $\sigma_R^2$ for different values
of the lead coupling $\Omega_L$ (Fig.~\ref{fig:decay-rates}).
The decay width $\Lambda$ has been obtained by choosing, for each curve, a suitable time $\hat{t}$ 
right before the occurrence of finite-size effects and computing $\Lambda=-(\hbar/\hat{t})\log P(\hat{t})$.
From our numerical results we observe that the agreement between
reduced and extended model is completely lost when the dephasing strength
$\sigma_R^2$ equals the lead coupling $\Omega_L$ (see vertical dashed
lines in Fig.~\ref{fig:decay-rates}). {We have also computed $\Lambda$
in presence of a finite (but small compared to $\Omega_L$) 
coupling between the sites of the ring. The results, shown as
crosses in Fig.~\ref{fig:decay-rates}, confirm that a presence of
a small $\Omega_R$ does not change the global picture. }

We can then conclude that it is possible to include in an additive way the effect of dephasing 
and of the presence of a probability-absorbing channel if the intensity $\sigma_R^2$ of dephasing 
is smaller than the channel bandwidth.
Moreover, the critical dephasing for which the additive approximation
breaks down is proportional  to the inter-site coupling $\Omega_L$ in the lead.

%%%%%%%%%%%%%%%%%%%%%%%%%%%%%%%%%%%%%%%%%%%%%%%%%%%%%%%%%%%%
\section{Conclusions}\label{sec:conclusions}

Transport in quantum networks is often affected by the interplay of
different environments. Typically, the effect of probability-absorbing baths is taken into
account by adding non-Hermitian terms to the Hamiltonian of the
system, while other environments are modeled by appropriate super-operators included in the master equation.
The action of both types of environment usually enters the master
equation in an additive way. The basis of the assumption is that the two baths affect the 
system independently and do not interfere with each other.

The aim of this work was to investigate the limit of validity of such an assumption
by means of both analytical derivations and numerical simulations.
To this end, we analyzed a simple quantum network (which is a paradigmatic model for transport phenomena) under the influence of two different baths: a probability-absorbing environment, represented by a lead, and a dephasing environment, modeled by white-noise fluctuations of the site energies.

Our analysis has shown that the additive approximation is valid when the strength of the time-dependent 
energy fluctuations is small compared to the bandwidth of the probability-absorbing bath.
In this case, the master equation for the open quantum system can be obtained by adding the contributions generated by a
non-Hermitian Hamiltonian, encoding the loss of probability, and by a Haken--Strobl super-operator, representing the dephasing bath.
In the opposite regime, the breakdown of the additive assumption leads to distinctive features such as the counterintuitive suppression of the decay when the coupling to the dephasing bath is increased.

To generalize our results, we stress that the large bandwidth approximation corresponds to the case in which the coupling with the probability-absorbing bath does not depend on the energy of the system states. In our case, such energy dependence is strong only for energies close to the band-edge of the lead, so that  if the fluctuations induced by the dephasing environment bring the system energies close to the lead band-edges,  the additive approximation breaks down. The general principle we can extract for different physical situations is the following: when the non-Hermitian description of a probability-absorbing bath is valid in the absence of other environments, it will remain valid even in the presence other environments when the fluctuations induced by the latter are so small that the energy dependence of the coupling with the probability-absorbing bath can be neglected. In particular, this will always be the case if the fluctuations  induced by the other environments are comparable with the system bandwidth. It could remain valid in principle for much larger strength of the fluctuations, like in the case studied in this paper, where we have shown that the relevant energy scale is the bandwidth of the lead and not the bandwidth of the ring system.

The main applicative implication of our investigation regards engineered systems for photon sensing or light harvesting. In proposals for such devices, see Ref.~\cite{superabsorb}, the acceptor system is modeled as a semi-infinite lead as we did in our paper. 
Thus, our results have a direct impact on the modeling and on the design of devices where the couplings between the different components can be tuned to optimize the performance of the system.

In excitonic transport in natural light-harvesting complexes, dephasing is often modelled by independent random fluctuations of site energies as we did here. Moreover, non-Hermitian terms are used to model excitation loss by trapping or recombination. Indeed, in natural light-harvesting complexes, there are two main ways in which the excitation can leave the system: (i) by recombination and photon emission; (ii) by trapping and charge separation in reaction centers. 
They constitute two independent probability-absorbing baths. 
As for the electromagnetic environment, its bandwidth is clearly very large since the photon can have any energy, moreover thermal
fluctuations ($\approx 200$ $\mathrm{cm}^{-1}$) are only a tiny fraction of the excitation energy of the single molecule ($\approx 10^4$ $\mathrm{cm}^{-1}$) and they are comparable with the system bandwidth.  
For this reason, even if a more quantitative analysis should be carried out, our results supports the widespread use of an effective non-Hermitian Hamiltonian entering additively in the master equation, following for example Ref.~\cite{mukameldeph}.
Regarding the loss of charge carriers by trapping in a reaction center, the actual physical processes involved are more complicated. 
All we can say is that care should be taken in modeling the interaction with the reaction centers and the wide-band condition should always be discussed on the basis of a more detailed analysis of each specific natural system.

%%%%%%%%%%%%%%
\section*{Acknowledgments}
G.G.G.~acknowledges support from the Okinawa Institute of Science and Technology Graduate University with subsidy funding from the Cabinet Office, Government of Japan, and from the Universit\`a Cattolica del Sacro Cuore through its research promotion activities.
G.S. has been supported by the DFG (SCHA 1646/3-1, GRK 1588, SFB 910).
G.L.C.~acknowledges useful discussion with F. Borgonovi. 

%%%%%%%%%%%%%%%%%%%%
\appendix

\section{Derivation of Haken--Strobl equation}\label{sec:HS}

Here we will consider the Haken--Strobl master equation for the average density matrix which describes a system in the presence of stochastic fluctuations of the site energies (see~\cite{HS}). 
In this section we introduce a simple way to derive the Haken-Strobl master equation, by using It\^o's stochastic calculus \cite{evans}. 
The starting point is a \emph{stochastic Schr\"odinger equation} in the standard form (see \cite{capitolo}):
\begin{equation}\label{eq:scrstoc}
\begin{aligned}
d\vc{\psi}(t) =\mbox{}& \Bigg( -\frac{\iu}{\hbar} H(t) - \frac{1}{2}\sum_{j}R_{j}^{*}(t)R_{j}(t)\Bigg)\vc{\psi}(t)dt\\
 &\mbox{}+ \sum_{j}R_{j}(t)\vc{\psi}(t)dW_{j}(t)\,, \\
\vc{\psi}(0) =\mbox{}& \vc{\psi}_{0}\,.
\end{aligned}
\end{equation}
Alongside the deterministic Hamiltonian term $-(\iu /\hbar)H(t)\vc{\psi}(t)dt$, we have a number of white-noise potentials $R_j(t)dW_j(t)$, and 
the term $1/2\sum_{j}R_{j}^{*}(t)R_{j}(t)$, necessary to conserve the total probability.
{Each $dW_j(t)$ denotes the stochastic differential of an independent Wiener process and is characterized by a variance proportional to the time increment, namely $\langle dW_j^2\rangle\propto dt$.}
Note that Eq.~\eqref{eq:scrstoc} is a linear It\^o's stochastic differential equation of the form
$d\vc{\psi} = \vc{F}dt + \vc{G}d\vc{W}\,$.

To model a system with $N$ sites (in the single-excitation approximation) with independent fluctuations of the site energies we assume that the operators $R_{j}(t)$, $j=1,\ldots,N$, are constant in time and have the form:
\begin{equation}\label{R equation}
R^{\alpha}_{\phantom{\alpha}\beta j } = \ -\frac{\iu}{{\sqrt{\hbar}}}\sigma_{j}\delta^{\alpha}_{\phantom{\alpha}\beta j}\ ,
\end{equation}
where $\sigma_j>0$ indicates the intensity of the noise on site $j$, and $\delta_{ijk}$ is the 3-index Kronecker symbol.
We have then with the following identifications:
\begin{equation}\label{G and F equations}
\begin{aligned}
F^{\alpha} =& \Bigg( -\frac{\iu}{\hbar} H^{\alpha}_{\phantom{\alpha}\beta} - \frac{1}{2\hbar}\sum_{j}\sigma^{2}_{j}\delta^{\alpha}_{\phantom{\alpha}\beta j}\Bigg)\psi^{\beta}\,, \\
G^{\alpha}_{\phantom{\alpha}j} =& -\frac{\iu}{{\sqrt{\hbar}}}\sigma_j\delta^{\alpha}_{\phantom{\alpha}\beta j}\psi^{\beta}\,.
\end{aligned}
\end{equation}  
From now on we assume summation over Greek repeated indices, while in the case of Latin indices the sum, if present, will be always explicitly written. In components on the site-basis Eq.~(\ref{eq:scrstoc}) reads
\begin{equation}\label{Schroedinger Stochastic eq in components}
\begin{aligned}
d\psi^{\alpha} =\mbox{}& \left(-\frac{\iu}{\hbar} H^{\alpha}_{\phantom{\alpha}\beta}- \frac{1}{2\hbar}\sum_{j}\sigma^{2}_{j}\delta^{\alpha}_{\phantom{\alpha}\beta j}\right)\psi^{\beta}dt \\
&\mbox{}- \frac{\iu}{{\sqrt{\hbar}}} \sum_{j}\sigma_{j}\delta^{\alpha}_{\phantom{\alpha}\beta j}dW_{j}\psi^{\beta}\,.
\end{aligned}
\end{equation}
We observe that the white-noise terms
\begin{equation}
\vt{V_j(t)}{\alpha}{\beta}\equiv\sigma_j\vt{\delta}{\alpha}{\beta j}\,dW_j(t)\,,\quad\text{for}\quad j=1,\ldots,N\,,
\end{equation}
represent the random fluctuations of the energy of each site ($j$) with intensity given by $\sigma_j^2dt$.

We recall that It\^o's product formula for the stochastic differential of two processes $X$ and $Y$ such that
\begin{equation}
dX=F_1dt+G_1dW\,,\qquad dY=F_2dt+G_2dW\,,
\end{equation}
reads
\begin{equation}
\label{eq:itoformula}
d(XY)=YdX+XdY+G_1G_2dt\,.  
\end{equation}

By applying It\^o's rule \eqref{eq:itoformula}, we can obtain from Eq.~\eqref{eq:scrstoc} the Quantum Stochastic Master Equation (QSME), governing the evolution of the random density matrix $\ket{\vc{\psi}}\bra{\vc{\psi}}$. In components, the QSME reads
\begin{equation}\label{eq:QSME}
\begin{aligned}
d\Big(\psi^{\gamma}&\psi^{*}_{\lambda}\Big) =\mbox{} -\frac{\iu}{\hbar}\left(H^{\gamma}_{\phantom{\gamma}\beta}\psi^{\beta}\psi^{*}_{\lambda}-\psi^{\gamma}\psi^{*}_{\beta}H^{* \beta}_{\phantom{* \beta}\lambda} \right)dt\\
&\mbox{}- \frac{\iu}{{\sqrt{\hbar}}} \sum_{j}\sigma_{j}\Big(\delta^{\gamma}_{\phantom{\gamma}\beta j}\psi^{\beta}\psi^{*}_{\lambda}\,dW_{j}-\psi^{\gamma}\psi^{*}_{\beta}\delta^{\beta}_{\phantom{\beta}\lambda j}\,dW_{j}\Big) \\
&\mbox{} - \frac{1}{2\hbar}\sum_{j}\sigma^{2}_{j}\Big(  \delta^{\beta}_{\phantom{\beta}\lambda j}\psi^{\gamma}\psi^{*}_{\beta} 
+ \delta^{\gamma}_{\phantom{\gamma}\beta j}\psi^{\beta}\psi^{*}_{\lambda} \Big) dt\\
&\mbox{} + \frac{1}{2\hbar}\sum_{j}\sigma^{2}_{j}\Big(\delta^{\gamma}_{\phantom{\gamma}\tau j}\delta^{\rho}_{\phantom{\rho}\lambda j}\psi^{\tau}\psi^{*}_{\rho} + \delta^{\gamma}_{\phantom{\gamma}\tau j}\delta^{\rho}_{\phantom{\rho} \lambda j}\psi^{* \tau}\psi_{\rho}\Big) dt\,.
\end{aligned}
\end{equation}
By taking the expected value of the QSME \eqref{eq:QSME}, recalling that terms proportional to $dW_j$ have zero mean, we obtain the following 
equation for  
$\rho=\langle\ket{\vc{\psi}}\bra{\vc{\psi}}\rangle$:
\begin{equation}\label{Haken-Strobl equation}
\begin{aligned}
d\langle(&\psi^{\gamma}\psi^{*}_{\lambda})\rangle = -\frac{\iu}{\hbar}\Big\langle H^{\gamma}_{\phantom{\gamma}\beta}\psi^{\beta}\psi^{*}_{\lambda}-\psi^{\gamma}\psi^{*}_{\beta}H^{* \beta}_{\phantom{* \beta}\lambda} \Big\rangle dt\\
&\mbox{} - \frac{1}{2\hbar}\Big\langle\sum_{j}\sigma^{2}_{j}\Big( \delta^{\beta}_{\phantom{\beta}\lambda j}\psi^{\gamma}\psi^{*}_{\beta} 
+ \delta^{\gamma}_{\phantom{\gamma}\beta j}\psi^{\beta}\psi^{*}_{\lambda}\Big)\Big\rangle dt\\
&\mbox{} + \frac{1}{2\hbar}\Big\langle\sum_{j}\sigma^{2}_{j}\Big(\delta^{\gamma}_{\phantom{\gamma}\tau j}\delta^{\rho}_{\phantom{\rho}\lambda j}\psi^{\tau}\psi^{*}_{\rho} + \delta^{\gamma}_{\phantom{\gamma}\tau j}\delta^{\rho}_{\phantom{\rho} \lambda j}\psi^{* \tau}\psi_{\rho}\Big)\Big\rangle dt\,.
\end{aligned}
\end{equation}
The foregoing equation corresponds to the Haken--Strobl equation \cite{HS}, and can be rearranged in the more familiar form
\begin{equation}\label{eq:HSrl}
\frac{d\rho^{j}_{\phantom{i}k}}{dt} = -\frac{\iu}{\hbar} \left(H\rho - \rho H^{\dagger}\right)^{j}_{\phantom{j}k} - (1 - \delta^{j}_{\phantom{j}k})\left(\frac{\sigma_{j}^{2} + \sigma_{k}^{2}}{2\hbar}\right)\rho^{j}_{\phantom{j}k}\,.
\end{equation}

We emphasize that no assumption is necessary on the Hermitian nature of the Hamiltonian.
The foregoing result can be applied to the modeling of noise in the high-temperature limit for both the extended and reduced systems considered in the main text.  

\section{Haken--Strobl in Lindblad form}\label{app:LinHS}

With reference to the notation of Sect.~\ref{sec:model} and \ref{sec:reduction} above, we can express the Haken--Strobl master equation in terms of the projectors on single-excitation states $\ket{\alpha_i}$, defined in Eq.~\eqref{eq:basis}. 

The equation reads
\begin{multline}
\dot{\rho}(t)=-\frac{\iu}{\hbar}[H_R+H_L+H_{RL},\rho]\\
+\sum_{i=1}^{N_R+N_L}\frac{\sigma^2_{ii}}{\hbar}\left(\ket{\alpha_i}\bra{\alpha_i}\rho\ket{\alpha_i}\bra{\alpha_i}-\frac{1}{2}\{\ket{\alpha_i}\bra{\alpha_i},\rho\}\right)\,,
\end{multline}
where $\sigma^2_{ii}$ is the intensity of noise on the $i$-th state, i.e., $\sigma_{ii:i\le N_R} = \sigma_R$, and $\sigma_{ii:i>N_R} = \sigma_L = 0$, compare Eq.~\eqref{eq:HS-rl}.
Evaluating matrix elements of this equation recovers the original dephasing dissipator~\eqref{eq:rfull}.

\end{document}